\documentclass[twocolumn,notitlepage,nofootinbib]{revtex4-2}
\usepackage{amsmath,amssymb,bm,graphicx,hyperref}
\allowdisplaybreaks[1]

\renewcommand\d{\partial}
\newcommand\+{\dagger}
\newcommand\<{\langle}
\renewcommand\>{\rangle}
\newcommand\eps{\varepsilon}
\newcommand\ep{\eps_p}
\newcommand\eq{\eps_q}
\newcommand\eqq{\eps_{q'}}
\newcommand\Z{\mathbb{Z}}
\newcommand\C{\mathcal{C}}

\renewcommand\H{\mathcal{H}}
\newcommand\N{\mathcal{N}}
\renewcommand\P{\mathcal{P}}
\renewcommand\S{\mathcal{S}}
\DeclareMathOperator\sgn{sgn}

\let\Im\relax\DeclareMathOperator\Im{Im}

\begin{document}

\title{Bulk viscosity of dual Bose and Fermi gases in one dimension}

\author{Tomohiro Tanaka}
\author{Yusuke Nishida}
\affiliation{Department of Physics, Tokyo Institute of Technology,
Ookayama, Meguro, Tokyo 152-8551, Japan}

\date{June 2022}

\begin{abstract}
One-dimensional Bose and Fermi gases with contact interactions are known to exhibit the weak-strong duality, where the equilibrium thermodynamic properties of one system at weak coupling are identical to those of the other system at strong coupling.
Here, we show that such duality extends beyond the thermodynamics to the frequency-dependent complex bulk viscosity, which is provided by the contact-contact response function.
In particular, we confirm that the bulk viscosities of the Bose and Fermi gases agree in the high-temperature limit, where the systematic expansion in terms of fugacity is available at arbitrary coupling.
We also compute their bulk viscosities perturbatively in the weak-coupling limit at arbitrary temperature, which via the duality serve as those of the Fermi and Bose gases in the strong-coupling limit.
\end{abstract}

\maketitle

\section{Introduction}
A one-dimensional gas of bosons with a contact interaction is known as the Lieb-Liniger model~\cite{Lieb:1963a}, which has been a paradigmatic model in low-dimensional quantum many-body physics.
It is not just a mathematical toy but has been realized experimentally with ultracold atoms~\cite{Bloch:2008,Cazalilla:2011}.
The Lieb-Liniger model is exactly solvable with the Bethe ansatz, from which the ground-state energy and the excitation spectrum were obtained~\cite{Lieb:1963a,Lieb:1963b}.
Furthermore, the thermodynamic Bethe ansatz developed by Yang and Yang allows us to determine the equilibrium thermodynamic properties of the system at finite temperature ($T$) and chemical potential ($\mu$)~\cite{Yang:1969}, although static and dynamic correlation functions remain more challenging to compute in general~\cite{Caux:2006,Caux:2007}.

Another fascinating aspect of the Lieb-Liniger model lies in the fact that its dual system composed of fermions can be constructed via Girardeau's Bose-Fermi mapping, so that all the energy eigenvalues are identical between the two systems~\cite{Girardeau:1960}.
In particular, it is the weak-strong duality, where one system at weak coupling corresponds to the other system at strong coupling, and the dual Fermi system is known as the Cheon-Shigehara model~\cite{Cheon:1999}.
Although such duality applies to the equilibrium thermodynamic properties of the two systems, it does not apply to general correlation functions with an exception of the structure factor provided by the density-density response function~\cite{Brand:2005,Cherny:2006}.
If the weak-strong duality is established for other correlation functions, it should be highly valuable because their strong coupling regime can be accessed with the perturbation theory of the dual system.

\begin{figure}[b]
\includegraphics[width=\columnwidth]{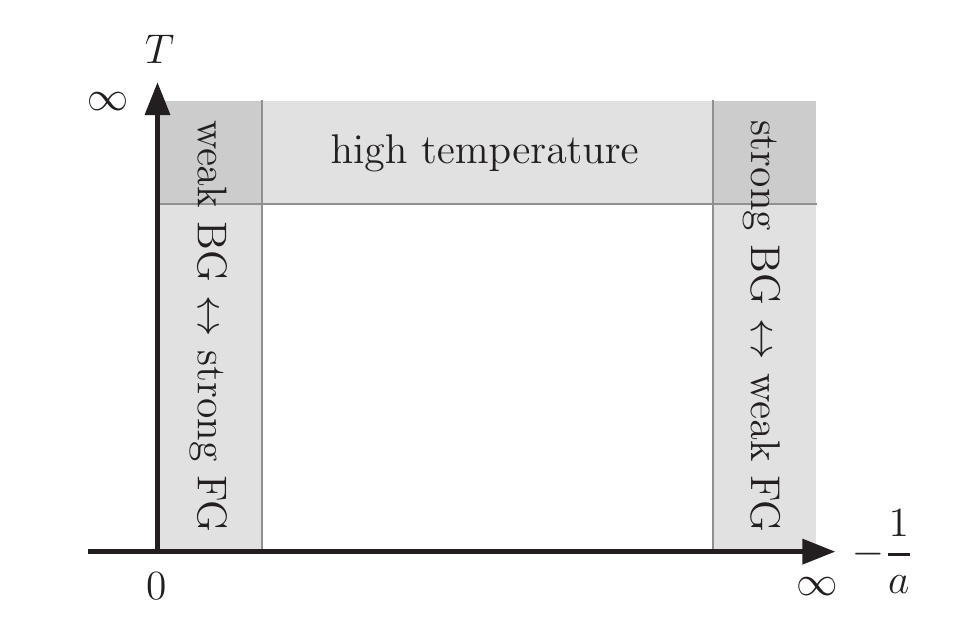}
\caption{\label{fig:duality}
Bulk viscosities of Bose and Fermi gases are evaluated in the high-temperature limit as well as in the weak-coupling limit, which corresponds to $a\to-\infty$ for bosons (BG) and $a\to-0$ for fermions (FG).
The system is thermodynamically unstable at $a>0$.}
\end{figure}

The purpose of this Letter is to extend the Bose-Fermi duality further to the frequency-dependent complex bulk viscosity, which has been subjected to active study for Fermi gases in higher dimensions as a measure of conformality breaking~\cite{Son:2007,Taylor:2010,Enss:2011,Hofmann:2011,Goldberger:2012,Taylor:2012,Dusling:2013,Chafin:2013,Martinez:2017,Fujii:2018,Nishida:2019,Enss:2019,Hofmann:2020,Fujii:2020,Maki:2020}.
To this end, we apply Girardeau's Bose-Fermi mapping to its Kubo formula, showing that the bulk viscosities of one-dimensional Bose and Fermi gases with contact interactions are identical at the same scattering length ($a$).
We also compute the bulk viscosities in the high-temperature limit at arbitrary coupling as well as in the weak-coupling limit at arbitrary temperature both for bosons and for fermions.
The two results in the high-temperature limit are useful to confirm the Bose-Fermi duality explicitly, whereas those in the weak-coupling limit are applicable to the Fermi and Bose gases in the strong-coupling limit, as indicated in Fig.~\ref{fig:duality}.

We will set $\hbar=k_B=1$ throughout this Letter and the bosonic and fermionic frequencies in the Matsubara formalism are denoted by $p_0=2\pi n/\beta$ and $p'_0=2\pi(n+1/2)/\beta$, respectively, for $n\in\Z$ and $\beta=1/T$.
Also, an integration over wave number or momentum is denoted by $\int_p\equiv\int_{-\infty}^\infty dp/(2\pi)$ for the sake of brevity, whereas the same definitions as in Ref.~\cite{Fujii:2020} are employed for the response function and Kubo's canonical correlation function [see Eqs.~(10) and (11) therein].

\section{Bose-Fermi duality}\label{sec:duality}
According to the linear-response theory~\cite{Mori:1962,Luttinger:1964,Bradlyn:2012}, the complex bulk viscosity at frequency $\omega$ is microscopically provided by
\begin{align}\label{eq:kubo}
\zeta(\omega) = \beta K_{\pi\pi}(w)
+ \frac\N{iw}\left(\frac{\d p}{\d\N}\right)_{\S/\N}
\end{align}
with the substitution of $w\in\mathbb{C}\to\omega+i0^+$ on the right-hand side~\cite{Fujii:2020}.
Here, $K_{\pi\pi}(w)$ is Kubo's canonical correlation function of the stress operator (scalar in one dimension) at zero wave number, whereas $p$, $\N$, and $\S$ are the pressure, the number density, and the entropy density, respectively.
In particular, its zero-frequency limit $\zeta\equiv\zeta(0)$ is referred to as the static bulk viscosity, which is real and corresponds to the bulk viscosity in hydrodynamics.

We first apply the above Kubo formula to the Lieb-Liniger model described by
\begin{align}
\hat\H = \frac{\d_x\hat\phi^\+\d_x\hat\phi}{2m}
+ \frac{g_B}{2}\hat\phi^\+\hat\phi^\+\hat\phi\hat\phi,
\end{align}
where $\hat\phi$ is the bosonic field operator and the coupling constant is related to the scattering length via $g_B=-2/(ma)>0$.
From the momentum continuity equation $\d_t\hat\P+\d_x\hat\pi=0$ for $\hat\P=[\hat\phi^\+(\d_x\hat\phi)-(\d_x\hat\phi^\+)\hat\phi]/(2i)$, the stress operator is found to be
\begin{align}\label{eq:stress_bose}
\hat\pi = 2\hat\H + \frac{\hat\C}{ma} - \frac{\d_x^2(\hat\phi^\+\hat\phi)}{4m}
\end{align}
with $\hat\C\equiv\hat\phi^\+\hat\phi^\+\hat\phi\hat\phi$ being the so-called contact density operator~\cite{Sekino:2021}.
Then, by substituting Eq.~(\ref{eq:stress_bose}) into Eq.~(\ref{eq:kubo}) and by employing the thermodynamic identities such as $dp=\S dT+\N d\mu-(\C/ma^2)da$ as detailed in Ref.~\cite{Fujii:2020}, we obtain
\begin{align}\label{eq:bulk_bose}
\zeta(\omega) = \frac1{iw}\frac{R_{\C\C}(w)}{(ma)^2}
+ \frac1{iw}\frac1m\left(\frac{\d\C}{\d a}\right)_{\N,\S},
\end{align}
where $\C=\<\hat\C\>$ is the contact density and $R_{\C\C}(w)$ is the response function of the contact density operator at zero wave number.

Let us next turn to the Cheon-Shigehara model described by
\begin{align}
\hat\H = \frac{\d_x\hat\psi^\+\d_x\hat\psi}{2m}
+ \frac{g_F}{2}(\d_x\hat\psi^\+)\hat\psi^\+\hat\psi(\d_x\hat\psi),
\end{align}
where $\hat\psi$ is the fermionic field operator and the coupling constant is related to the scattering length via
\begin{align}\label{eq:coupling_fermi}
\frac1{g_F} = -\frac{m\Lambda}{\pi} + \frac{m}{2a} < 0
\end{align}
with $\Lambda$ being the momentum cutoff for regularization~\cite{Cui:2016,Sekino:2021}.
Although the three-body interaction term $\sim|(g_F\hat\psi\d_x\hat\psi)\hat\psi|^2$ is necessary for the complete correspondence to the Lieb-Liniger model beyond the two-body level~\cite{Sekino:2018a,Sekino:2021}, it can be omitted for our analysis below within the two-body level because such a scale invariant term does not contribute to the stress operator other than in the Hamiltonian density~\cite{footnote}.
From the momentum continuity equation $\d_t\hat\P+\d_x\hat\pi=0$ for $\hat\P=[\hat\psi^\+(\d_x\hat\psi)-(\d_x\hat\psi^\+)\hat\psi]/(2i)$, the stress operator is found to be
\begin{align}\label{eq:stress_fermi}
\hat\pi = 2\hat\H + \frac{\hat\C}{ma} - \frac{\d_x^2(\hat\psi^\+\hat\psi)}{4m}
\end{align}
with $\hat\C\equiv(mg_F/2)^2(\d_x\hat\psi^\+)\hat\psi^\+\hat\psi(\d_x\hat\psi)$ being the contact density operator.
Again, by substituting Eq.~(\ref{eq:stress_fermi}) into Eq.~(\ref{eq:kubo}) and by employing the thermodynamic identities~\cite{Fujii:2020}, we obtain
\begin{align}\label{eq:bulk_fermi}
\zeta(\omega) = \frac1{iw}\frac{R_{\C\C}(w)}{(ma)^2}
+ \frac1{iw}\frac1m\left(\frac{\d\C}{\d a}\right)_{\N,\S},
\end{align}
which is expressed by the contact density operator in exactly the same form as Eq.~(\ref{eq:bulk_bose}).

We now recall that the Lieb-Liniger model and the Cheon-Shigehara model share the common energy eigenvalues and thus the common equilibrium thermodynamic properties at the same scattering length~\cite{Girardeau:1960,Cheon:1999}.
If the matrix elements of the contact density operator with respect to any energy eigenstates are also common between the two systems, their contact-contact response functions prove to be identical as seen in the Lehmann representation,
\begin{align}
R_{\C\C}(w) = -\frac1{ZL}\sum_{n,n'}
\frac{e^{-\beta E_n} - e^{-\beta E_{n'}}}{w+E_n-E_{n'}}|\<n|\hat{C}|n'\>|^2,
\end{align}
where $Z=\sum_ne^{-\beta E_n}$, $L=\int dx$, and $\hat{C}=\int dx\,\hat\C(x)$ are introduced.
Indeed, the resulting matrix elements in first quantization read as
\begin{align}\label{eq:contact_bose}
& \<n|\hat\C(x)|n'\> = N(N-1)\int dx_3\cdots dx_N \notag\\
&\quad \times \Phi_n^*(x,x,x_3,\dots,x_N)\Phi_{n'}(x,x,x_3,\dots,x_N)
\end{align}
for $N$ bosons and
\begin{align}\label{eq:contact_fermi}
& \<n|\hat\C(x)|n'\> = N(N-1)\int dx_3\cdots dx_N \notag\\
&\quad \times \Psi_n^*(x,x+0^+,x_3,\dots,x_N)\Psi_{n'}(x,x+0^+,x_3,\dots,x_N)
\end{align}
for $N$ fermions.
Because their wave functions are related to each other via Girardeau's Bose-Fermi mapping~\cite{Girardeau:1960},
\begin{align}
\Psi_n(x_1,x_2,\dots,x_N)
= \Biggl[\prod_{i<j}\sgn(x_i-x_j)\Biggr]\Phi_n(x_1,x_2,\dots,x_N),
\end{align}
Eqs.~(\ref{eq:contact_bose}) and (\ref{eq:contact_fermi}) are identical with the mapping factor squared being unity.
Consequently, the Bose-Fermi duality is established for the frequency-dependent complex bulk viscosity provided by Eqs.~(\ref{eq:bulk_bose}) and (\ref{eq:bulk_fermi}) equivalently, which constitutes the main outcome of this Letter.
In particular, it is the weak-strong duality, where $a\to-\infty$ corresponds to weakly interacting bosons but to strongly interacting fermions and $a\to-0$ corresponds to strongly interacting bosons but to weakly interacting fermions, as indicated in Fig.~\ref{fig:duality}.

\section{Bulk viscosity of a Bose gas}\label{sec:bose}
\begin{figure}[b]
\includegraphics[width=\columnwidth]{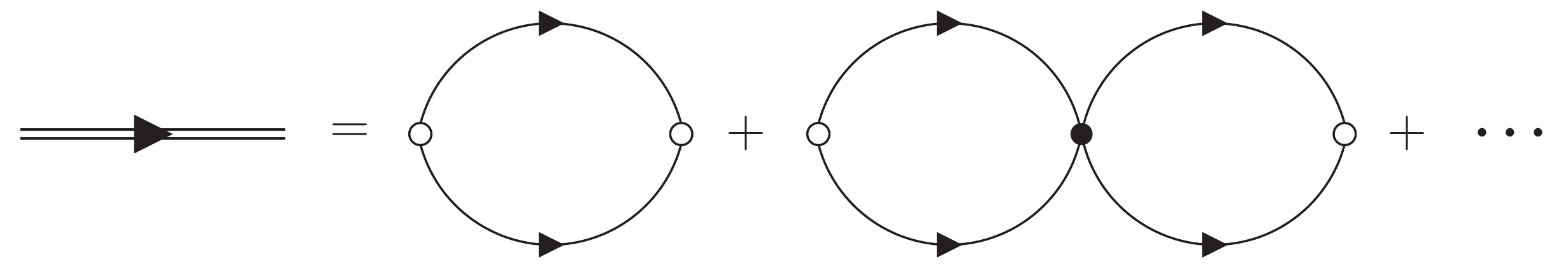}\\\bigskip
\includegraphics[width=0.75\columnwidth]{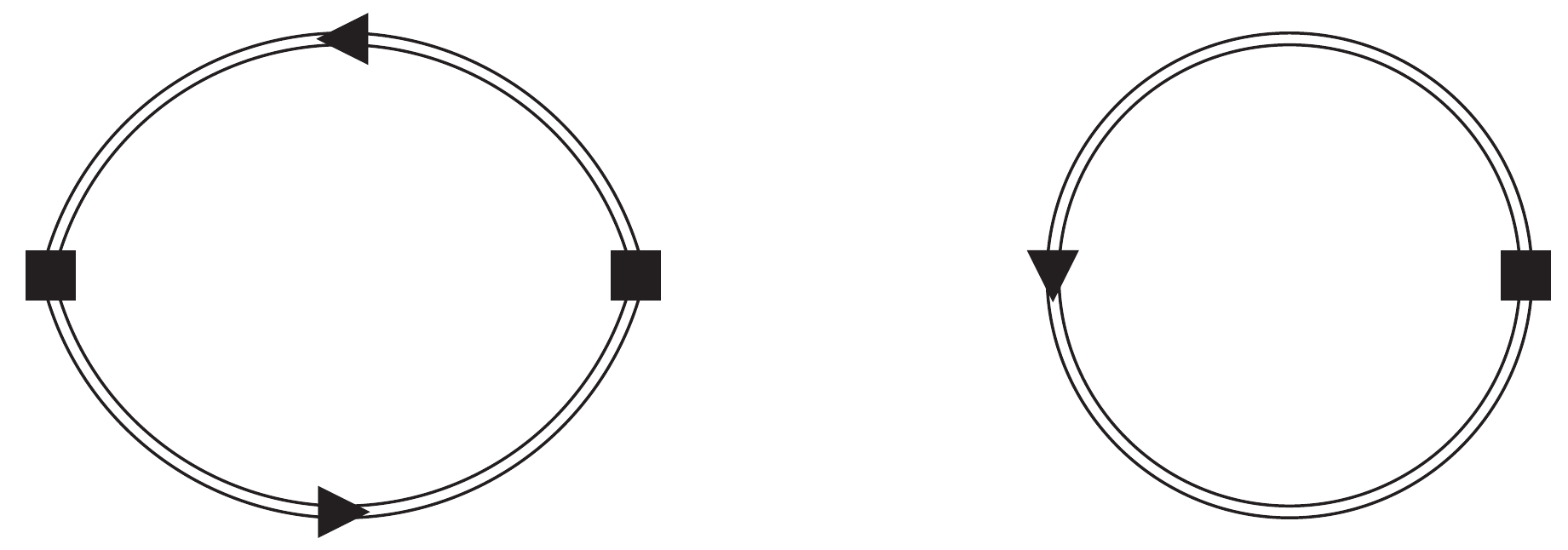}
\caption{\label{fig:contact}
Diagrammatic representation of (upper) the pair propagators in Eqs.~(\ref{eq:propagator_bose}) and (\ref{eq:propagator_fermi}), (lower left) the contact-contact response function in Eq.~(\ref{eq:response}), and (lower right) the contact density in Eq.~(\ref{eq:contact}).
The single and double lines represent the single-particle and pair propagators, respectively, whereas the filled dot is the interaction vertex carrying the coupling constant $-g_{B,F}/2$, the open dot carries just unity for bosons but $mg_F/2$ for fermions, and the square is to insert the contact density operator.}
\end{figure}

We then evaluate the frequency-dependent complex bulk viscosity in Eq.~(\ref{eq:bulk_bose}) for the Lieb-Liniger model both in the high-temperature limit and in the weak-coupling limit, where systematic expansions in terms of small parameters are available.
For the sake of dealing with the two cases as coherently as possible, it is convenient to follow Ref.~\cite{Fujii:2020} by introducing the pair propagator in the medium,
\begin{align}\label{eq:propagator_bose}
& D_B(ip_0,p) = -\sum_{n=0}^\infty\left(-\frac{g_B}{2}\right)^n \notag\\
&\quad \times \left[\frac2\beta\sum_{q_0}\int_q\,G(ip_0-iq_0,p-q)G(iq_0,q)\right]^{n+1},
\end{align}
whose diagrammatic representation is depicted in Fig.~\ref{fig:contact}.
Here, $G(iq_0,q)=1/(iq_0-\eq)$ with $\eq=q^2/(2m)-\mu$ is the single-particle propagator and the Matsubara frequency summation leads to
\begin{align}
\frac1{D_B(ip_0,p)} = \frac1{ma}
+ \left[2\int_q\,\frac{1+f_B(\eps_{p/2-q})+f_B(\eps_{p/2+q})}
{ip_0-\eps_{p/2-q}-\eps_{p/2+q}}\right]^{-1}
\end{align}
with $f_B(\eps)=1/(e^{\beta\eps}-1)$ being the Bose-Einstein distribution function.

The contact-contact response function in terms of the pair propagator is provided by
\begin{align}\label{eq:response}
R_{\C\C}(ik_0) = \frac1\beta\sum_{p_0}\int_p\,D_B(ik_0+ip_0,p)D_B(ip_0,p),
\end{align}
whose diagrammatic representation is also depicted in Fig.~\ref{fig:contact}.
The Matsubara frequency summation can be performed by employing the spectral representation of the pair propagator,
\begin{align}
D_B(ip_0,p) = \int_{-\infty}^\infty\!\frac{dE}{\pi}\,\frac{\Im[D_B(E-i0^+,p)]}{ip_0-E},
\end{align}
so that we obtain
\begin{align}\label{eq:continuation}
& R_{\C\C}(w) = -\int_p\iint_{-\infty}^\infty\!\frac{dE}{\pi}\frac{dE'}{\pi}\,
[f_B(E)-f_B(E')] \notag\\
&\quad \times \frac{\Im[D_B(E-i0^+,p)]\Im[D_B(E'-i0^+,p)]}{w+E-E'}
\end{align}
under the analytic continuation of $ik_0\to w$.
Similarly, the contact density in terms of the pair propagator is depicted in Fig.~\ref{fig:contact} and reads as
\begin{align}\label{eq:contact}
\C = -\frac1\beta\sum_{p_0}\int_p\,e^{ip_00^+}D_B(ip_0,p).
\end{align}
Its partial derivative with respect to $a$ is provided by
\begin{align}
\left(\frac{\d\C}{\d a}\right)_{\beta,\mu}
= -\frac1{ma^2}\frac1\beta\sum_{p_0}\int_p\,[D_B(ip_0,p)]^2,
\end{align}
which together with Eq.~(\ref{eq:response}) leads to
\begin{align}\label{eq:derivative}
\frac{R_{\C\C}(w)}{(ma)^2} + \frac1m\left(\frac{\d\C}{\d a}\right)_{\beta,\mu}
= \frac{R_{\C\C}(w) - R_{\C\C}(0)}{(ma)^2}.
\end{align}

\begin{figure}[t]
\includegraphics[width=0.9\columnwidth]{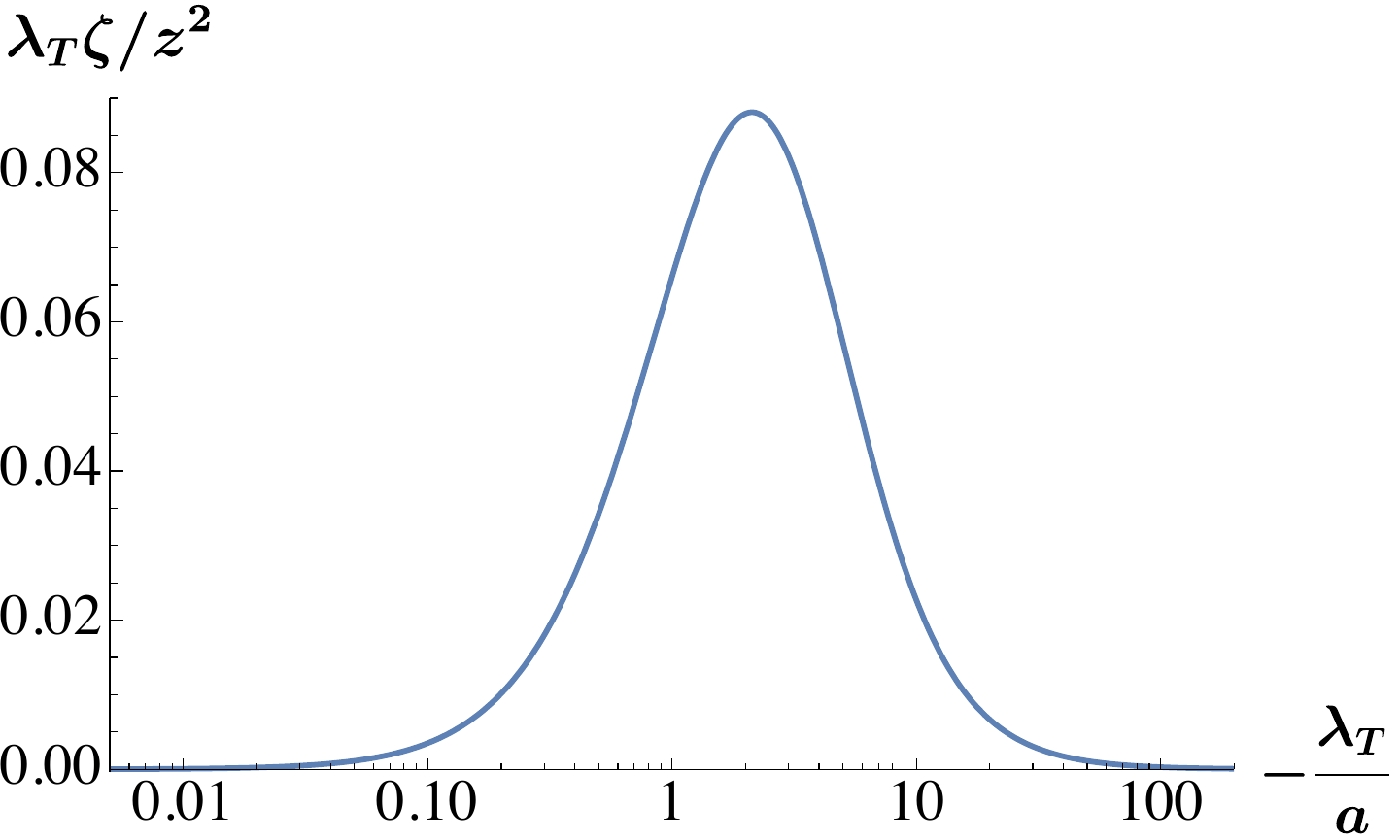}
\caption{\label{fig:high-T}
Static bulk viscosity of the Lieb-Liniger model and the Cheon-Shigehara model in the high-temperature limit $z\to0$ from Eq.~(\ref{eq:high-T}) as a function of the inverse scattering length.
The number density is provided by $\N=z/\lambda_T+O(z^2)$.}
\end{figure}

We now focus on the high-temperature limit at fixed number density, where the fugacity $z=e^{\beta\mu}\to0$ serves as a small parameter for the quantum virial expansion~\cite{Liu:2013}.
Its dependence in Eq.~(\ref{eq:continuation}) can be exposed by changing the integration variables to $\eps^{(\prime)}=E^{(\prime)}-p^2/(4m)+2\mu$ and then by expanding the distribution functions as $f_B(\eps-n\mu)=z^ne^{-\beta\eps}+O(z^{n+1})$.
Because the pair propagator in Eq.~(\ref{eq:propagator_bose}) fully incorporates two-body physics, the contact-contact response function in Eq.~(\ref{eq:response}) is actually exact up to $O(z^2)$ where only two-body physics is relevant~\cite{Nishida:2019,Fujii:2020}.
Furthermore, because of $(\d\C/\d a)_{\N,\S}=(\d\C/\d a)_{\beta,\mu}+O(z^3)$, Eqs.~(\ref{eq:continuation}) and (\ref{eq:derivative}) substituted into Eq.~(\ref{eq:bulk_bose}) lead to
\begin{align}\label{eq:high-T}
& \zeta(\omega) = \frac{\sqrt2\,z^2}{(ma)^2\lambda_T}
\iint_{-\infty}^\infty\!\frac{d\eps}{\pi}\frac{d\eps'}{\pi}\,
\frac{e^{-\beta\eps} - e^{-\beta\eps'}}{\eps-\eps'} \notag\\
&\quad \times \frac{\Im[D_0(\eps-i0^+)]\Im[D_0(\eps'-i0^+)]}
{i(\omega+\eps-\eps'+i0^+)} + O(z^3),
\end{align}
where $\lambda_T=\sqrt{2\pi\beta/m}$ is the thermal de Broglie wavelength and $D_0(\eps)\equiv D_B(\eps+p^2/4m-2\mu,p)|_{z\to0}=m/(1/a-\sqrt{-m\eps})$ is the pair propagator in the center-of-mass frame in the vacuum.
The resulting formula constitutes the frequency-dependent complex bulk viscosity of the Lieb-Liniger model in the high-temperature limit to the lowest order in fugacity.
In particular, the static bulk viscosity at $O(z^2)$ varies nonmonotonically under the inverse scattering length as plotted in Fig.~\ref{fig:high-T}.
Here, it is found to vanish analytically
\begin{align}\label{eq:high-T_fermi}
\zeta \to \frac{2\sqrt2\,z^2a^2}{\lambda_T^3}
\end{align}
at $a\to-0$ but nonanalytically
\begin{align}\label{eq:high-T_bose}
\zeta \to \frac{z^2\lambda_T}{\sqrt2\pi^2a^2}
\ln\!\left(\frac{2\pi a^2}{e^{1+\gamma}\lambda_T^2}\right)
\end{align}
at $a\to-\infty$, which closely resembles the static bulk viscosity of a two-component Fermi gas in three dimensions~\cite{Nishida:2019,Enss:2019,Hofmann:2020}.

\begin{figure}[b]
\includegraphics[width=0.9\columnwidth]{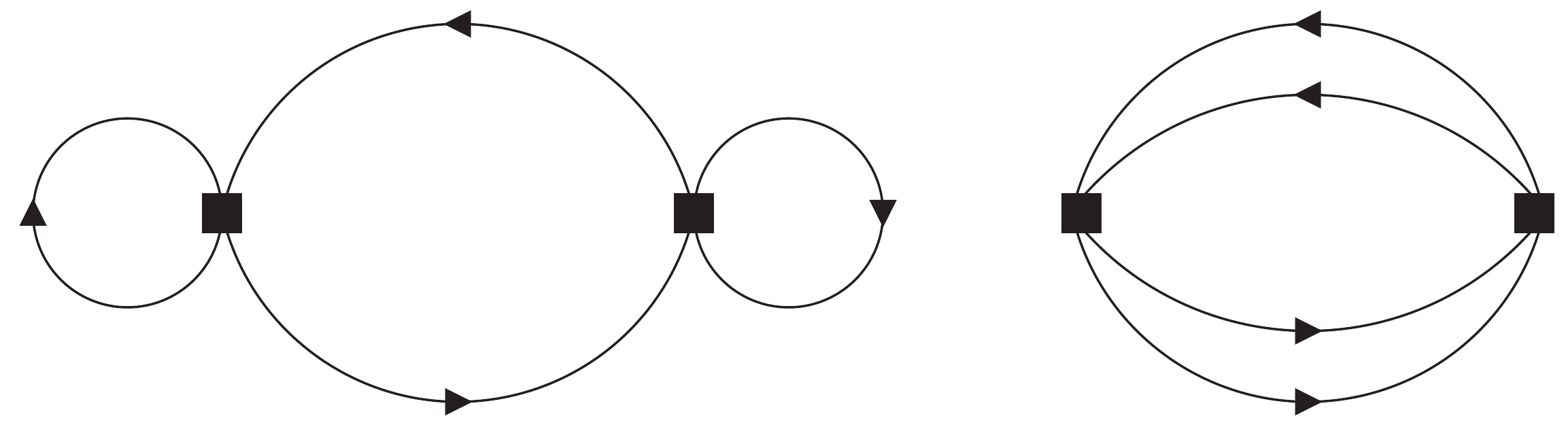}
\caption{\label{fig:perturbation}
Diagrams potentially contributing to the contact-contact response function in the weak-coupling limit.}
\end{figure}

\begin{figure}[t]
\includegraphics[width=0.9\columnwidth]{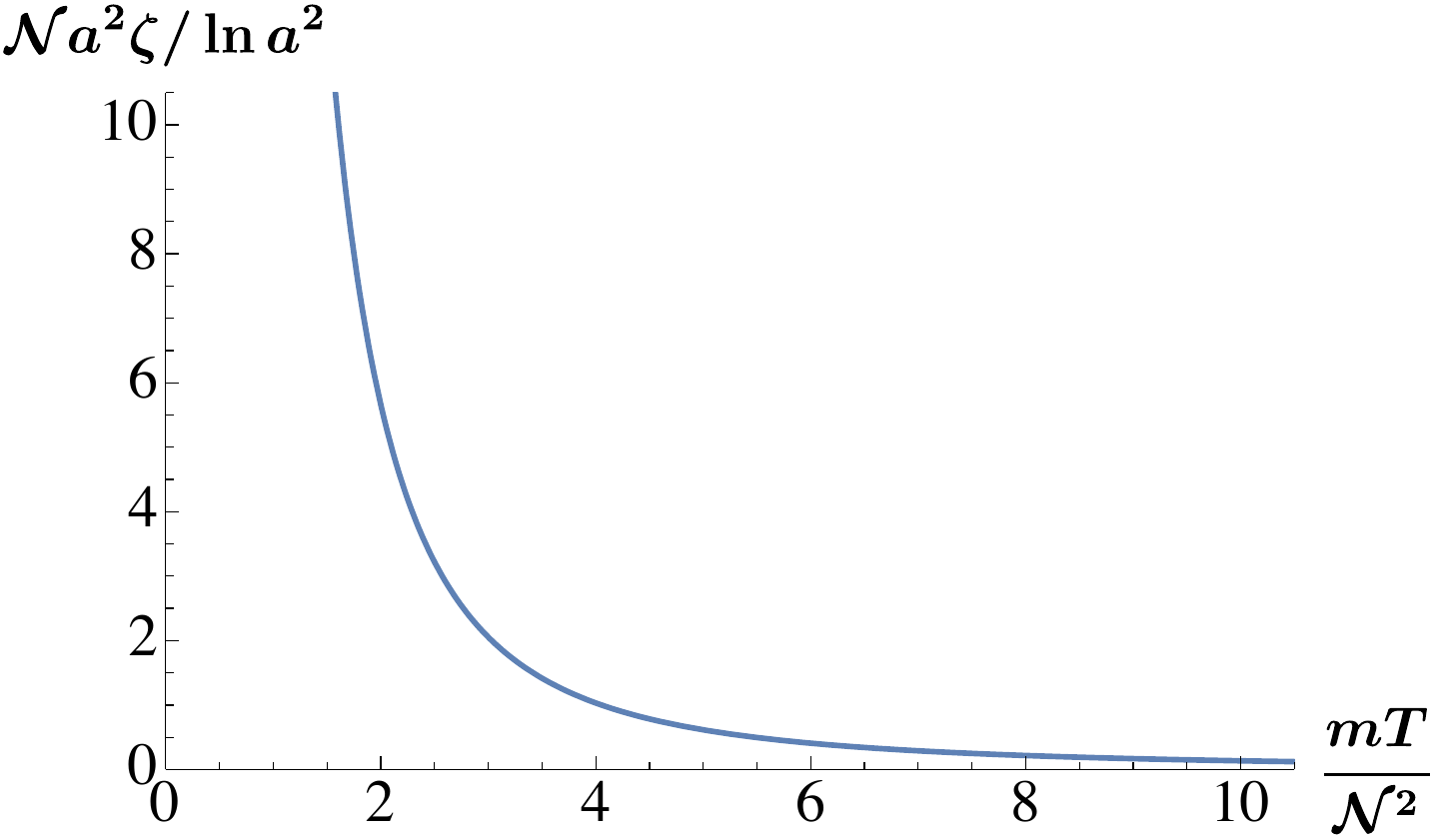}
\caption{\label{fig:weak_bose}
Static bulk viscosity of the Lieb-Liniger model in the weak-coupling limit $a\to-\infty$, corresponding to the Cheon-Shigehara model in the strong-coupling limit, from Eq.~(\ref{eq:weak_bose}) as a function of the temperature.
The number density is provided by $\N=\int_pf_B(\ep)+O(a^{-1})$.}
\end{figure}

On the other hand, there exist only two diagrams depicted in Fig.~\ref{fig:perturbation} that potentially contribute to the contact-contact response function in the weak-coupling limit $a\to-\infty$.
However, the left diagram therein proves to vanish at zero wave number, whereas the remaining right diagram is already incorporated in Fig.~\ref{fig:contact} as the lowest-order diagram.
Therefore, the contact-contact response function in the weak-coupling limit is obtained simply by expanding Eq.~(\ref{eq:continuation}) to the lowest order in coupling.
Furthermore, because of $(\d\C/\d a)_{\N,\S}=(\d\C/\d a)_{\beta,\mu}+O(a^{-3})$, Eqs.~(\ref{eq:continuation}) and (\ref{eq:derivative}) substituted into Eq.~(\ref{eq:bulk_bose}) lead to
\begin{align}\label{eq:weak_bose}
& \zeta(\omega) = \left(\frac2{ma}\right)^2\int_{p,q,q'}
\frac{f_B(\eps_{p/2-q}+\eps_{p/2+q}) - f_B(q\to q')}{2\eq-2\eqq} \notag\\
&\quad \times \frac{[1+f_B(\eps_{p/2-q})+f_B(\eps_{p/2+q})]
[\,q\to q'\,]}{i(\omega+2\eq-2\eqq+i0^+)} + O(a^{-3}),
\end{align}
which constitutes the frequency-dependent complex bulk viscosity of the Lieb-Liniger model in the weak-coupling limit.
We note that the resulting formula at zero frequency is logarithmically divergent at $q\sim q'\sim0$, which is to be cut off by $1/|a|$ after its resummation as seen in the high-temperature limit.
Therefore, the static bulk viscosity to the lowest order in coupling is actually $O[(\ln a^2)/a^2]$, which varies monotonically under the temperature as plotted in Fig.~\ref{fig:weak_bose}.
Here, it is found to diverge as
\begin{align}
\zeta \to \frac{5\,\N^7\ln a^2}{\pi(mT)^4a^2}
\end{align}
at $T\to0$ and then decrease toward
\begin{align}
\zeta \to \frac{2\,\N^2\ln a^2}{\sqrt\pi(mT)^{3/2}a^2}
\end{align}
at $T\to\infty$ in agreement with the high-temperature limit in Eq.~(\ref{eq:high-T_bose}).

\section{Bulk viscosity of a Fermi gas}\label{sec:fermi}
Let us turn to the frequency-dependent complex bulk viscosity in Eq.~(\ref{eq:bulk_fermi}) for the Cheon-Shigehara model, which can be evaluated systematically both in the high-temperature limit and in the weak-coupling limit in parallel with the Lieb-Liniger model.
First, the pair propagator in the medium is provided by
\begin{align}\label{eq:propagator_fermi}
& D_F(ip_0,p) = -\left(\frac{mg_F}{2}\right)^2
\sum_{n=0}^\infty\left(-\frac{g_F}{2}\right)^n \notag\\
&\quad \times \left[\frac2\beta\sum_{q'_0}\int_q\,
(q-p/2)^2G(ip_0-iq'_0,p-q)G(iq'_0,q)\right]^{n+1},
\end{align}
whose diagrammatic representation is depicted in Fig.~\ref{fig:contact}.
Then, the Matsubara frequency summation leads to
\begin{align}
& \frac1{D_F(ip_0,p)} = \frac1{ma} \notag\\
&\quad - \frac2m\int_q\left[1 + \frac{q^2}{m}
\frac{1-f_F(\eps_{p/2-q})-f_F(\eps_{p/2+q})}{ip_0-\eps_{p/2-q}-\eps_{p/2+q}}\right],
\end{align}
where $f_F(\eps)=1/(e^{\beta\eps}+1)$ is the Fermi-Dirac distribution function and the regularization is applied under Eq.~(\ref{eq:coupling_fermi}) with $\Lambda\to\infty$.

The contact-contact response function and the contact density in terms of the above pair propagator are depicted in Fig.~\ref{fig:contact}, so that all the results presented in Eqs.~(\ref{eq:response})--(\ref{eq:derivative}) are valid just by replacing $D_B$ with $D_F$.
Because of $D_F(\eps+p^2/4m-2\mu,p)|_{z\to0}=m/(1/a-\sqrt{-m\eps})=D_0(\eps)$ in the high-temperature limit, the frequency-dependent complex bulk viscosity of the Cheon-Shigehara model to the lowest order in fugacity is provided by exactly the same formula as Eq.~(\ref{eq:high-T}) for the Lieb-Liniger model.
This is indeed expected and confirms the extended Bose-Fermi duality established by this Letter.

\begin{figure}[t]
\includegraphics[width=0.9\columnwidth]{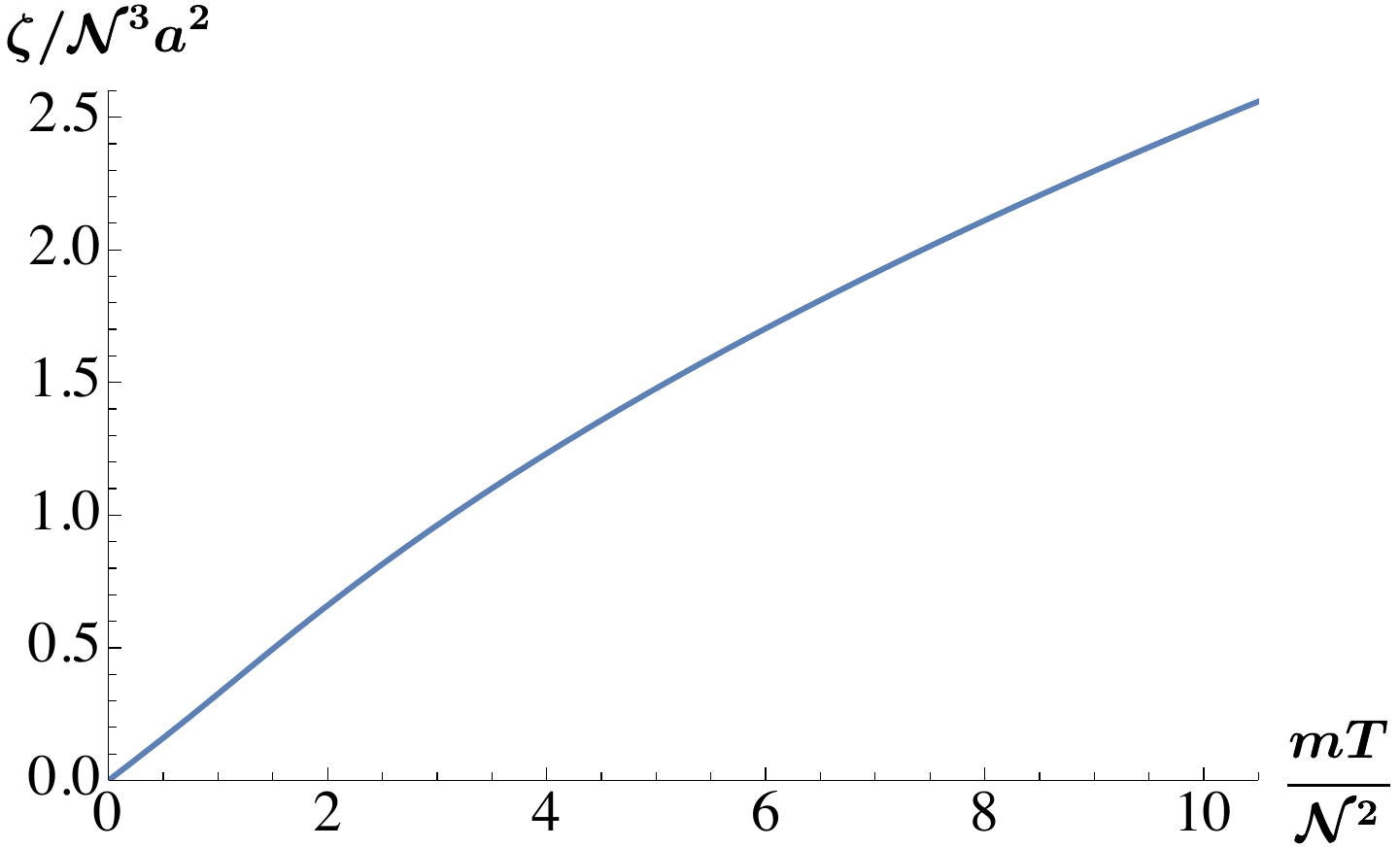}
\caption{\label{fig:weak_fermi}
Static bulk viscosity of the Cheon-Shigehara model in the weak-coupling limit $a\to-0$, corresponding to the Lieb-Liniger model in the strong-coupling limit, from Eq.~(\ref{eq:weak_fermi}) as a function of the temperature.
The number density is provided by $\N=\int_pf_F(\ep)+O(a)$.}
\end{figure}

Similarly, only the right diagram in Fig.~\ref{fig:perturbation} contributes to the contact-contact response function in the weak-coupling limit $a\to-0$, which is already incorporated in Fig.~\ref{fig:contact} as the lowest-order diagram.
Therefore, by expanding Eq.~(\ref{eq:continuation}) for $D_F$ instead of $D_B$ to the lowest order in coupling, we now obtain
\begin{align}\label{eq:weak_fermi}
& \zeta(\omega) = \left(\frac{2a}{m}\right)^2\int_{p,q,q'}
\frac{f_B(\eps_{p/2-q}+\eps_{p/2+q}) - f_B(q\to q')}{2\eq-2\eqq} \notag\\
&\quad \times q^2q'^2\frac{[1-f_F(\eps_{p/2-q})-f_F(\eps_{p/2+q})]
[\,q\to q'\,]}{i(\omega+2\eq-2\eqq+i0^+)} + O(a^3),
\end{align}
which constitutes the frequency-dependent complex bulk viscosity of the Cheon-Shigehara model in the weak-coupling limit.
Thanks to the extended Bose-Fermi duality, the resulting formula is also applicable to the Lieb-Liniger model in the strong-coupling limit, whereas Eq.~(\ref{eq:weak_bose}) is applicable to the Cheon-Shigehara model in the strong-coupling limit, both of which are otherwise inaccessible.
In particular, the static bulk viscosity at $O(a^2)$ varies monotonically under the temperature as plotted in Fig.~\ref{fig:weak_fermi}, where it is found to vanish as
\begin{align}
\zeta \to \frac{mT\N a^2}{\pi}
\end{align}
at $T\to0$ and then increase toward
\begin{align}
\zeta \to \frac{2(mT)^{1/2}\N^2a^2}{\sqrt\pi}
\end{align}
at $T\to\infty$ in agreement with the high-temperature limit in Eq.~(\ref{eq:high-T_fermi}).

\section{Summary and prospects}\label{sec:summary}
In summary, we showed that the weak-strong duality between one-dimensional Bose and Fermi gases with contact interactions extends beyond the thermodynamics to the frequency-dependent complex bulk viscosity.
This was achieved by applying Girardeau's Bose-Fermi mapping to its Kubo formula expressed in terms of the contact-contact response function.
We also evaluated their bulk viscosities in the high-temperature, weak-coupling, and strong-coupling limits (shaded regions in Fig.~\ref{fig:duality}), whose formulas are provided by Eqs.~(\ref{eq:high-T}), (\ref{eq:weak_bose}), and (\ref{eq:weak_fermi}) with their static limits presented in Figs.~\ref{fig:high-T}, \ref{fig:weak_bose}, and \ref{fig:weak_fermi}.
Although the static bulk viscosity was found to be finite at its lowest orders in all the three limits, a Drude peak divergent at zero frequency proves to appear at higher orders, which will be reported elsewhere~\cite{Nishida:preprint}.

Away from such limits where systematic expansions in terms of small parameters are unavailable, we plan to compute the frequency-dependent complex bulk viscosity numerically with the aid of the Bethe-ansatz solution along the line of Refs.~\cite{Caux:2006,Caux:2007}.
To this end, the sum rule and the high-frequency tail,
\begin{align}
\int_{-\infty}^\infty\!\frac{d\omega}{\pi}\,\zeta(\omega)
&= -\frac1m\left(\frac{\d\C}{\d a}\right)_{\N,\S}, \\
\lim_{|\omega|\to\infty}\zeta(\omega) &= \frac{\C}{|m\omega|^{3/2}a^2}
- \frac{i}{m\omega}\left(\frac{\d\C}{\d a}\right)_{\N,\S},
\end{align}
valid at arbitrary temperature, density, and scattering length will be useful to benchmark the numerics.

The resulting bulk viscosity may be relevant to the hydrodynamics of a one-dimensional gas of bosons realized with ultracold atoms~\cite{Bloch:2008,Cazalilla:2011}, where the integrability is inevitably broken at long timescales, for example, by an emergent three-body interaction~\cite{Mazets:2010,Tanaka:preprint}.
Furthermore, the frequency-dependent complex bulk viscosity can be extracted experimentally by measuring the contact, energy, or entropy density under the periodic modulation of the scattering length~\cite{Fujii:2018,Nishida:2019}.
Hopefully, our findings in this Letter shed new light on nonequilibrium dynamics of ultracold atoms in one dimension.

\acknowledgments
The authors thank Yuta Sekino for valuable discussions as well as Jeff Maki and Shizhong Zhang for correspondences regarding their related work~\cite{Maki:preprint}.
This work was supported by JSPS KAKENHI Grants No.\ JP18H05405 and No.\ JP21K03384.

\end{document}